\documentclass[twocolumn,eqsecnum,aps]{revtex4}
\usepackage[dvips]{epsfig,color}
\usepackage{amsmath,amssymb}
\def\PRL{Phys. Rev. Lett. }
\def\PRB{Phys. Rev. B }
\def\JPSJ{J. Phys. Soc. Jap. }
\def\PLB{Phys. Lett. B }
\def\NPB{Nucl. Phys. B }
\def\IJMPB{Int. J. Mod. Phys. B }
\def\bbox#1{\boldsymbol#1}
\def\ZeeEne{\varDelta_{\txZ}}
\def\txZ{\text{Z}}
\def\a{{\cal A}}
\def\DiracU{\Phi_{\text{D}}}
\def\LLL{lowest Landau level}
\def\dxdy{d^{2}xd^{2}y}
\def\upA{\uparrow}
\def\dnA{\downarrow}
\def\UnitZ{|g^{*}\mu_{B}B|}
\def\eff{\text{eff}}
\def\txf{\text{f}}
\def\txb{\text{b}}
\def\txg{\text{g}}
\def\DSAS{\varDelta_{\text{SAS}}}
\def\sky{\text{sky}}
\def\spin{\text{spin}}
\def\ppin{\text{ppin}}
\def\ipin{\text{ipin}}
\def\txS{\text{S}}
\def\txA{\text{A}}
\def\H{{\cal H}}
\def\frS{T}
\def\frbS{\bbox{T}}
\def\sS{P}
\def\sbS{\bbox{P}}
\def\F{\varphi }
\def\ddq{\!{d^{2}q\frac{1}{2\pi}}}
\begin{document}
\title{Interlayer Exchange Interactions, SU(4) Soft Waves and Skyrmions\\
in Bilayer Quantum Hall Ferromagnets
}
\author{Z. F. Ezawa and K. Hasebe}
\affiliation{
Department of Physics, Tohoku University, Sendai, 980-8578 Japan
}
\date{\today}
\begin{abstract}
The Coulomb exchange interaction is the driving force for quantum coherence in 
quantum Hall systems.  We construct a microscopic Landau-site Hamiltonian for 
the exchange interaction in bilayer quantum Hall ferromagnets, which is 
characterized by the SU(4) isospin structure.  By taking a continuous limit, 
the Hamiltonian gives rise to the SU(4) nonlinear sigma model in the 
von-Neumann-lattice formulation.  The ground-state energy is evaluated at 
filling factors $\nu =1,2,3,4$.  It is shown at $\nu =1$ that there are 3 
independent soft waves, where only one soft wave is responsible for the 
coherent tunneling of electrons between the two layers.  It is also shown at 
$\nu =1$ that there are 3 independent skyrmion states apart from the 
translational degree of freedom.  They are CP$^{3}$ skyrmions enjoying the 
spin-charge entanglement confined within the \LLL.
\end{abstract}
\maketitle

\section{Introduction}

Quantum coherence is a new aspect of quantum Hall (QH) systems.  
Electron spins are polarized spontaneously due to the exchange Coulomb 
interaction rather than compulsively by the Zeeman effect.  Skyrmions arise as 
coherent excitations\cite{Sondhi93B}, which have been observed 
experimentally\cite{Barrett95L,Aifer96L,Schmeller95L} as quasiparticles.  
Intriguingly, an interlayer coherence may develop spontaneously between the 
two layers and lead to Josephson-like phenomena in bilayer QH (BLQH)  
systems\cite{Ezawa92IJMPB}.  Recent experimental results\cite{Spielman00L} on 
tunnelling current may well be interpreted as the dc-Josephson 
current\cite{Fogler01L} though still controversial\cite{Stern01L}.  We expect even the 
SU(4) quantum coherence due to the spin and layer degrees of 
freedom\cite{Ezawa99L}.  The driving force of quantum coherence is the interlayer 
exchange interaction\cite{Moon95B,Ezawa97B}.  The exchange Coulomb interaction has 
also been argued to create a new phase, the canted 
antiferromagnet\cite{Zheng97L,Sarma97L,Demler99L}, in the BLQH system at the 
filling factor $\nu =2$.

In this paper we analyze the exchange Coulomb interaction to explore 
the SU(4) coherence in BLQH ferromagnets.  We are concerned about electrons 
confined to the \LLL, where the electron position is solely specified by the 
guiding center $(X,Y)$ obeying
\begin{equation}
[X,Y] = -i\ell _{B}^{2} .
\label{IntroGuidi}
\end{equation}
This brings in the noncommutative W$_{\infty }$ algebra\cite{Girvin84B} as the basic 
symmetry of the QH system.  It implies that the electron position cannot be 
localized to a point within the \LLL, and hence the system cannot be described 
by local field theory.  We construct an effective field theory to describe 
physics whose scale is larger than the magnetic length $\ell _{B}\equiv \sqrt {\hbar /eB}$.

The effective Hamiltonian governing the SU(2) coherence has been 
derived\cite{Moon95B,Ezawa97B} by making a derivative expansion of the Coulomb 
energy of spin or pseudospin textures at $\nu =1$, where the spin stiffness $J$ 
is explicitly calculated as
\begin{equation}
J = \frac{1} {16\sqrt{2\pi}}\frac{e^{2}}{4\pi\varepsilon \ell _{B}} .
\label{IntroSpinStiff}
\end{equation}
It is a straightforward but complicated task to generalize the SU(2) scheme to 
the SU(4) scheme, because the SU(4) extension of the W$_{\infty }$  algebra is 
considerably complicated than the SU(2) extension\cite{Ezawa97B}.

We overcome the problem by employing an alternative formulation.  
Namely, we construct a Landau-site Hamiltonian by expanding the electron field 
operator in terms of the one-body wave functions of electrons confined to the 
\LLL.  Then, the exchange Coulomb interaction emerges just as in ferromagnets.  
An effective Hamiltonian is derived by taking a continuum limit in the 
von-Neumann-lattice formulation, where we substitute the spin stiffness 
(\ref{IntroSpinStiff}) for the exchange integral.  This is a consistent procedure 
in the context of the SU(2) coherence.  We generalize this procedure to study 
the SU(4) coherence.  In the SU(4)-invariant limit the effective Hamiltonian 
is given by the SU(4) nonlinear sigma model 
\begin{equation}
H_{X}^{\eff} = 2J \sum _{a=1}^{15}\int d^{2}x [\partial _{k}\frS_{a}(\bbox{x})]^{2} ,
\label{IntroExchaSUFour}
\end{equation}
where $\frS_{a}(\bbox{x})$ is the isospin field normalized as $\sum _{a}\frS_{a}\frS_{a}=3/8$ at 
$\nu =1$.  The present approach allows us to analyze QH states at any filling 
factor.  For the sake of simplicity, we discuss only integer QH states, though  
fractional QH states are treated similarly in the framework of the composite 
fermion theory\cite{Jain89L}.

It is our main purpose to make a thorough investigation of soft waves 
and skyrmion excitations supported by the exchange Hamiltonian, though some of 
them have been known previously\cite{Ezawa99L,Ghosh01B}.  We examine carefully 
what are the dynamical fields in the BLQH system.  There are three degenerate 
soft waves in the SU(4)-invariant limit, among which only one soft wave is 
responsible for the coherent tunneling of electrons between the two layers.  
The soft modes are Goldstone modes associated with spontaneous breakdown of 
the SU(4) isospin symmetry.  Actually, the degeneracy is resolved by the 
Zeeman effect and the tunneling interaction.  Namely, the SU(4) symmetry is 
broken explicitly but softly by these interactions, and Goldstone modes turn 
into pseudo-Goldstone modes with gaps.  It is also shown at $\nu =1$ that there 
are three independent skyrmion states apart from the translational degree of 
freedom.  They are CP$^{3}$ skyrmions enjoying the spin-charge entanglement 
confined within the \LLL.

\section{Quantum Hall Ferromagnets}

To elucidate quantum coherence we start with monolayer QH systems.  
Electrons make cyclotron motions under perpendicular magnetic field $B$.  The 
number of flux quanta passing through the system is $N_{\Phi }\equiv BS/\DiracU$, where $S$ 
is the area and $\DiracU=2\pi \hbar /e$ is the flux quantum.  There are $N_{\Phi }$ electron 
states per one Landau level by neglecting the spin degree of freedom, each of 
which is associated with one flux quantum.  We call it the Landau site.  One 
Landau site occupies the area $S/N_{\Phi }=2\pi \ell _{B}^{2}$, and may accommodate two electrons 
with up and down spins.  The filling factor is $\nu =N/N_{\Phi }$ with $N$ the total 
number of electrons.  We are concerned about physics taking place in the \LLL.

The microscopic Hamiltonian is a sum of the Coulomb term and the 
Zeeman term,
\begin{align}
H_{\text{C}}&= {\frac{1} {2}}\int \dxdy V(\bbox{x}-\bbox{y})\rho (\bbox{x})\rho (\bbox{y}),  \label{CouloEnerg}\\
H_{\text{Z}}&= -{\frac{1} {2}}\ZeeEne \int d^{2}x \bigl[\rho ^{\upA }(\bbox{x})-\rho ^{\dnA }(\bbox{x})\bigr] ,  \label{ZeemaEnerg}
\end{align}
where $V(\bbox{x}-\bbox{y})=e^{2}/4\pi \varepsilon |\bbox{x}-\bbox{y}|$ is the Coulomb potential; $\rho ^{\sigma }(\bbox{x})=\psi ^{\sigma \dagger }(\bbox{x})\psi ^{\sigma }(\bbox{x})$ is 
the electron density with the spin index $\sigma =\upA ,\dnA $; $\rho (\bbox{x})=\rho ^{\upA }(\bbox{x})+\rho ^{\dnA }(\bbox{x})$; 
$\ZeeEne=\UnitZ$ is the Zeeman gap with $g^{*}$ the magnetic g-factor and 
$\mu _{B}$ the Bohr magneton.

We expand the electron field operator in terms of the one-body wave 
functions $\F_{i}(\bbox{x})$ in the \LLL,
\begin{equation}
\psi ^{\sigma }(\bbox{x}) \equiv  \sum _{i=1}^{N_{\Phi }} c_{\sigma }(i)\F_{i}(\bbox{x}) ,
\label{FieldExpan}
\end{equation}
where $c_{\sigma }(i)$ is the annihilation operator of the up-spin ($\sigma =\upA $) or down-spin 
($\sigma =\dnA $) electron at the Landau site $i$,
\begin{align}
\{c_{\sigma }(i),c_{\tau }^{\dagger }(j)\}&=\delta _{ij}\delta _{\sigma \tau },\notag\\ 
\{c_{\sigma }(i),c_{\tau }(j)\}&= \{c_{\sigma }^{\dagger }(i),c_{\tau }^{\dagger }(j)\} =0 .
\label{AntiCommuC}
\end{align}
As is well known\cite{Iso92PLB,Stone90B}, it is impossible to choose an 
orthonormal complete set of one-body wave functions $\varphi _{i}(\bbox{x})$ in the expansion 
(\ref{FieldExpan}).  Consequently, the electron field $\psi ^{\sigma }(\bbox{x})$ does not satisfy a 
standard canonical anticommutation relation, $\{\psi ^{\sigma }(\bbox{x}),\psi ^{\sigma \dagger }(\bbox{y})\}\not=\delta (\bbox{x}-\bbox{y})$, 
implying that an electron cannot be localized to a point within the \LLL.  
Nevertheless, roughly speaking, an electron can be localized into a Landau 
site, which has the area $2\pi \ell _{B}^{2}$.  Thus, it is reasonable that QH effects are 
described by the Landau-site Hamiltonian.

Substituting (\ref{FieldExpan}) into (\ref{CouloEnerg}), we derive the direct and 
exchange Coulomb energies,
\begin{subequations}
\begin{align}
H_{D} &= {\frac{1}{2}}\sum _{i,j} U_{ij}n(i)n(j),     \label{HamilDirec}\\
H_{X} &= 2 \sum _{\langle i,j\rangle }\sum _{\sigma ,\tau } J_{ij}c^{\dagger }_{\sigma }(i)c^{\dagger }_{\tau }(j)c_{\tau }(i)c_{\sigma }(j) , \label{HamilFerroPre}
\end{align}
\end{subequations}
where $n(i)\equiv \sum _{\sigma }c^{\dagger }_{\sigma }(i)c_{\sigma }(i)$ is the electron number at site $i$:  $U_{ij}$ and 
$J_{ij}$ are the direct and exchange integrals,
\begin{subequations}
\begin{align}
U_{ij}&= \int \!d^{2}xd^{2}y\; \F_{i}^{*}(\bbox{x})\F_{j}^{*}(\bbox{y})V(\bbox{x}-\bbox{y})\F_{i}(\bbox{x})\F_{j}(\bbox{y}),\\
J_{ij}&= {\frac{1}{2}}\int \!d^{2}xd^{2}y\; \F_{i}^{*}(\bbox{x})\F_{j}^{*}(\bbox{y})V(\bbox{x}-\bbox{y})\F_{i}(\bbox{y})\F_{j}(\bbox{x}).\label{ExchaInteg}
\end{align}
\end{subequations}
These integrals are convergent because the wave function $\F_{i}(\bbox{x})$ is 
"localized" within one Landau site $i$ with area $2\pi \ell _{B}^{2}$.   The sum $\sum _{\langle i,j\rangle }$ 
runs over all spin pairs ($i\not=j$) just once.

The spin $\bbox{S}(i)$ is defined at each site $i$ by
\begin{equation}
S_{a} = (c_{\upA }^{\dagger },c_{\dnA }^{\dagger }) {\frac{\tau _{a}}{2}} \begin{pmatrix}c_{\upA }\\ c_{\dnA } \end{pmatrix}
\label{ExchaPauli}
\end{equation}
with $\tau _{a}$ the Pauli matrix.  Using the algebraic relation
\begin{equation}
\sum _{\sigma ,\tau } c^{\dagger }_{\sigma }(i)c^{\dagger }_{\tau }(j)c_{\tau }(i)c_{\sigma }(j) = - 2\bbox{S}(i)\!\cdot\! \bbox{S}(j) -{\frac{1}{2}}n(i)n(j) ,
\label{ExchaCC}
\end{equation}
we rewrite the exchange term as
\begin{equation}
H_{X} = - 4\sum _{\langle i,j\rangle } J_{ij} \bbox{S}(i)\!\cdot\! \bbox{S}(j) - \sum _{\langle i,j\rangle } J_{ij} n(i)n(j) .
\label{HamilFerro}
\end{equation}
The Hamiltonian has the global O(3) symmetry:  It is invariant when all spins 
are rotated simultaneously. 

At $\nu =1$, in the absence of the Zeeman effect (\ref{ZeemaEnerg}), the spin 
direction is determined spontaneously to minimize the exchange energy.  The 
product $\bbox{S}(i)\!\cdot\! \bbox{S}(j)$ takes the maximum value, $\bbox{S}(i)\!\cdot\! \bbox{S}(j)=1/4$, when 
$\bbox{S}(i)=\bbox{S}(j)$.  Hence, provided $J_{ij}>0$, all spins are spontaneously polarized 
to minimize the exchange energy, where the direction of polarization is 
arbitrary:  $\bbox{S}(i)=\bbox{S}$ for all points $i$ but the direction of $\bbox{S}$ is arbitrary.  
Actually, the direction of the polarization is the $z$ axis due to the Zeeman 
effect however small it may be.  The exchange interaction contributes to the 
ground-state energy by
\begin{equation}
\langle H_{X}\rangle _{\txg} = -2\sum _{\langle i,j\rangle } J_{ij} = -N_{\Phi } \sum _{j} J_{ij} = - {\frac{1}{2}}\varepsilon _{X}N ,
\label{GrounEnerg}
\end{equation}
where
\begin{equation}
\varepsilon _{X} \equiv 2\sum _{j} J_{ij}
\end{equation}
with $i$ fixed arbitrarily; the sum runs over all sites for $J_{ij}\not=0$.  It 
is clear that the loss of the exchange energy is $\varepsilon _{X}$ when one electron is 
removed from filled Landau sites.  This is equal to the energy necessary to 
flip one spin in the system (\ref{HamilFerro}).  

At $\nu =2$ we obtain 
$\langle H_{X}\rangle _{\txg}=-\varepsilon _{X}N_{\Phi }=- {\frac{1}{2}}\varepsilon _{X}N$ since $\bbox{S}(i)=0$ and 
$n(i)=2$ in (\ref{HamilFerro}).
\begin{figure}[tbp]
\includegraphics*[width=85mm]{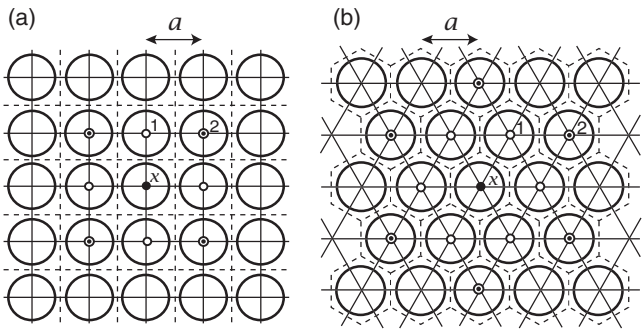}
\renewcommand{\baselinestretch}{1.0}
\caption{
An electron makes a cyclotron motion occupying an area $2\pi \ell _{B}^{2}$ 
and avoiding all others.  Spin-polarized electrons fill the \LLL\ at the 
filling factor $\nu =1$.  Their configuration is represented by a von 
Neumann lattice with the lattice point identified with the center of the 
cyclotron motion.  Lattice points in the nearest neighborhood of the point 
$\bbox{x}$ are designated by open circles numbered by 1.  Lattice points in 
the second nearest neighborhood are designated by double circles numbered by 
2.  A square lattice (a) and a triangular lattice (b) are examples of von 
Neumann lattices.
}
\label{CycloLatPS}
\end{figure}

\section{von Neumann Lattice}\label{SecNeumaLatti}

Because the QH system is robust against density fluctuations, the 
direct Coulomb term (\ref{HamilDirec}) is irrelevant as far as perturbative 
fluctuations are concerned.  We wish to analyze the exchange interaction 
(\ref{HamilFerro}).  As we have stated, the exchange integral (\ref{ExchaInteg}) is 
convergent.  Recall that we have expanded the electron field in terms of 
one-body wave functions $\F_{i}(\bbox{x})$ as in (\ref{FieldExpan}).  The index $i$ may 
represent the angular momentum in the symmetric gauge or the linear momentum 
in the Landau gauge.  When we evaluate the exchange integral either in the 
symmetric or Landau gauges, we find a large contribution from a spin pair 
$\langle i,j\rangle $ even if they are not in the nearest neighborhood of each other.  
Furthermore, it is not clear at all how the rotational and translational 
symmetries are recovered in these gauges when the continuum limit is taken:  
See Ref.\cite{Ghosh01B} for instance.

It is most convenient for us to use a set of one-body wave functions 
$\F_{i}(\bbox{x})$ in (\ref{FieldExpan}) so that the index $i$ runs over a lattice such as a 
square lattice or a triangular lattice [Fig.\ref{CycloLatPS}] with the lattice 
point being the center of the cyclotron motion.  We can construct such a 
lattice with the use of the coherent-state representation.  

We adopt the symmetric gauge, where $A_{x}= {\frac{1}{2}}By$ and $A_{y}
=-{\frac{1}{2}}Bx$.  
The angular momentum is given by $L=\hbar b^{\dagger }b$ in the \LLL\ with
\begin{align}
b \equiv  {\frac{1}{\sqrt {2}\ell _{B}}}(X - iY) = {\frac{1}{\sqrt {2}}}\biggl(z^{*} + {\frac{\partial}{\partial z}}\biggr), \notag\\
b^{\dagger }\equiv  {\frac{1}{\sqrt {2}\ell _{B}}}(X + iY) = {\frac{1}{\sqrt {2}}}\biggl(z - {\frac{\partial}{\partial z^{*}}}\biggr),
\label{NeumaB}
\end{align}
where $z=(x+iy)/2\ell _{B}$.  We introduce an eigenstate of the angular-momentum 
lowering operator $b$,
\begin{equation}
b|\beta \rangle  = \beta |\beta \rangle  .
\label{NeumaA}
\end{equation}
Because $b$ is an annihilation operator, the state $|\beta \rangle $ is a coherent state 
by definition, and is given by
\begin{equation}
|\beta \rangle  \equiv  e
^{\beta b^{\dagger }-\beta ^{*}b}|0\rangle  = e^{-|\beta |^{2}/2} e^{\beta b^{\dagger }}|0\rangle  ,
\label{NeumaC}
\end{equation}
where $|0\rangle $ is the angular-momentum zero state obeying $b|\beta \rangle =0$.  The wave 
function $\F_{\beta }(\bbox{x})=\langle \bbox{x}|\beta \rangle $ is calculated as
\begin{equation}
\F_{\beta }(\bbox{x})={\frac{1}{\sqrt {2\pi \ell _{B}^{2}}}}\exp\biggl(-\bigm|\!z-{\frac{1}{ \sqrt {2}}}\beta \bigm|^{2} +{\frac{i(y\beta _{\Re}+x\beta _{\Im})}{ \sqrt {2}\ell _{B}}}\biggr) ,
\label{NeumaWaveExpli}
\end{equation}
where $\beta =\beta _{\Re}+i\beta _{\Im}$.  It describes an electron localized around the point 
$z=\beta /\sqrt {2}$.

The coherent state has the minimum uncertainty subject to the 
Heisenberg uncertainty relation associated with the noncommutativity 
(\ref{IntroGuidi}) between the coordinates $X$ and $Y$.  The state $|\beta \rangle $ 
corresponds to the classical state describing a cyclotron motion around the 
point
\begin{equation}
x = \sqrt {2}\ell _{B}\beta _{\Re},\quad \quad  y = -\sqrt {2}\ell _{B}\beta _{\Im} ,
\end{equation}
as follows from (\ref{NeumaB}) and (\ref{NeumaA}).  Because $\beta $ is an arbitrary complex 
number, an electron may be localized around any point.  

We consider the QH state at $\nu =1$.  The system is filled up with 
electrons each of which occupies an area $2\pi \ell _{B}^{2}$.  It is reasonable to put 
electrons on a lattice with the unit cell area $2\pi \ell _{B}^{2}$.  Such a lattice is 
nothing but a von Neumann 
lattice\cite{vonNeumann55,Perelomov71,Bargmann71,Boon81,Imai90B}.  The states on a 
von Neumann lattice form a minimum complete set\cite{Perelomov71,Bargmann71} in 
the \LLL.  Thus, we may expand the electron field in terms of coherent states 
$\F_{i}(\bbox{x})$ as in (\ref{FieldExpan}), where $i$ runs over all lattice points.

We consider a square lattice for simplicity [Fig.\ref{CycloLatPS}(a)].  
Lattice points are given by $\beta _{mn}=\sqrt {\pi }(m+in)$, or
\begin{equation}
X_{m} = \sqrt {2\pi }\ell _{B}m,\quad \quad Y_{n} = -\sqrt {2\pi }\ell _{B}n,
\end{equation}
so that the unit cell area is $2\pi \ell _{B}^{2}$.  States are given by 
\begin{equation}
|X_{m},Y_{n}\rangle = \exp\bigl[-\pi (m^{2}+n^{2})\bigr]\exp\bigl[\sqrt {\pi }(m+in)b^{\dagger }\bigr]|0\rangle  .
\end{equation}
They are not orthogonal,
\begin{equation}
\langle X_{m'},Y_{n'}|X_{m},Y_{n}\rangle =\exp\biggl(-{\frac{\pi}{2}}\bigl[(m-m')^{2}+(n-n')^{2}\bigr]\biggr).
\end{equation}
The wave function (\ref{NeumaWaveExpli}) reads
\begin{align}
\F_{mn}(\bbox{x})=&\langle \bbox{x}|X_{m},Y_{n}\rangle  \notag\\
=& {\frac{1}{\sqrt {2\pi \ell _{B}^{2}}}}\exp\biggl(-\bigm|z-{\frac{1}{\sqrt {2}}}\beta _{mn}\bigm|^{2}\biggr) \notag\\
&\times \exp\biggl(\frac{i\sqrt{\pi}}{\sqrt{2}\ell _{B}}(ym+xn)\biggr) ,
\label{OneBodyNeuma}
\end{align}
which describes an electron localized around the lattice point 
$(x,y)=(X_{m},Y_{n})$.

\section{Continuum Limit}\label{SecContiLimit}

In the von-Neumann-lattice formulation it is straightforward to take 
the field-theoretical limit of the exchange energy (\ref{HamilFerro}), by letting 
the lattice spacing $a$ vanish just as in a lattice model for ferromagnets.  
The resulting Hamiltonian describes correctly physical phenomena whose typical 
size is much larger than the spacing $a$.  

We first analyze the nearest-neighbor terms, for which we set $J_{ij}\equiv J_{1}$.  
Let the lattice points be specified by lattice vectors $\bbox{a}^{\alpha }$ with $\sum _{\alpha }\bbox{a}^{\alpha }=0$.  We 
expand the spin product as
\begin{align}
\sum _{\langle i,j\rangle }\bbox{S}_{i}\!\cdot\! \bbox{S}_{j}&= {\frac{1}{ 2}}\sum _{\bbox{x}}\sum _{\alpha }\bbox{S}(\bbox{x})\!\cdot\! \bbox{S}(\bbox{x}+\bbox{a}^{\alpha }) \notag\\
&\simeq  {\frac{1}{2}}\sum _{\bbox{x}}\sum _{\alpha }\bigl[\bbox{S}(\bbox{x})^{2} - {\frac{1}{2}}a_{i}^{\alpha }a_{j}^{\alpha } \partial _{i}\bbox{S}(\bbox{x}){\!\cdot\! }\partial _{j}\bbox{S}(\bbox{x})\bigr],
\label{LattiExpan}
\end{align}
where a partial integration was made.  The exchange Hamiltonian (\ref{HamilFerro}) 
yields
\begin{equation}
H_{X}^{\eff} \simeq  J_{1}\biggl(\sum _{\alpha }a_{i}^{\alpha }a_{j}^{\alpha }\biggr) \sum _{\bbox{x}}\partial _{i}\bbox{S}(\bbox{x}){\cdot }\partial _{i}\bbox{S}(\bbox{x}) 
\label{FerroSpin}
\end{equation}
as the lowest order term in the derivative expansion.  The ground-state energy 
is given by (\ref{GrounEnerg}) with $\varepsilon _{X}=2J_{1}\sum _{\alpha }$.  We next analyze the second 
nearest-neighbor terms with the lattice vectors $\bbox{b}^{\beta }$, for which we set 
$J_{ij}\equiv J_{2}$.  We obtain the same formula as (\ref{FerroSpin}) with the replacement of 
$\bbox{a}^{\beta }$ by $\bbox{b}^{\beta }$.  Any lattice points can be treated in the same way.  

We explicitly consider a square lattice [Fig.\ref{CycloLatPS}(a)] as a 
simplest von Neumann lattice, where $\sum _{\alpha =1}^{4}a_{i}^{\alpha }a_{j}^{\alpha }=2a^{2}\delta _{ij}$ and $\sum _{\bbox{x}}=a^{-2}\int \!d^{2}x$.  
Hence, (\ref{FerroSpin}) amounts to the O(3) nonlinear sigma model 
\begin{equation}
H_{X}^{\eff} = 2J \int \!d^{2}x\; \partial _{i}\bbox{S}(\bbox{x}){\cdot }\partial _{i}\bbox{S}(\bbox{x}) ,
\label{FieldHamil}
\end{equation}
where $J=J_{1}$ and $h_{X}=8J_{1}$ in the ground-state energy (\ref{GrounEnerg}) for the 
nearest-neighbor terms.  It is easy to see that $J= J_{1}+2J_{2}+4J_{3}+ \cdots  $ and
$\varepsilon _{X}=8J_{1}+8J_{2}+8J_{3}+ \cdots  $ by taking into account all lattice points; the series 
would converge rapidly.  When we adopt another lattice such as the triangular 
lattice [Fig.\ref{CycloLatPS}(b)] and take the continuum limit, we reproduce 
the same effective Hamiltonian (\ref{FieldHamil}) together with the ground-state 
energy (\ref{GrounEnerg}) but with different definitions of $J$ and $\varepsilon _{X}$ in terms 
of the exchange integrals $J_{ij}$.

We determine the parameters $J$ and $\varepsilon _{X}$ as follows.  The effective 
Hamiltonian (\ref{FieldHamil}) was first proposed to study skyrmion 
excitations\cite{Sondhi93B}, where the spin stiffness $J$ was 
identified\cite{Kallin84B} with
\begin{equation}
J = {\frac{1}{16\sqrt {2\pi }}}{\frac{e^{2}}{4\pi \varepsilon \ell _{B}}}.
\label{SpinStiff}
\end{equation}
The formula has been verified by evaluating explicitly the energy of a spin 
texture\cite{Moon95B,Ezawa97B}.  We next estimate the parameter $\varepsilon _{X}$, by 
substituting the skyrmion configuration 
\begin{equation}
S_{x} = {\frac{\kappa x}{ r^{2}+\kappa ^{2}}},\quad  S_{y} =-{\frac{\kappa y}{ r^{2}+\kappa ^{2}}},\quad  S_{z} = {\frac{1}{2}}{\frac{r^{2}-\kappa ^{2}}{ r^{2}+\kappa ^{2}}}
\label{SkyrmSigma}
\end{equation}
into the nonlinear sigma model (\ref{FieldHamil}).  One skyrmion increases the 
exchange energy\cite{Sondhi93B} by
\begin{equation}
\langle H_{X}\rangle _{\sky} = 4\pi J ,
\label{SkyrmEnergX}
\end{equation}
which is independent of the skyrmion size $\kappa $ in (\ref{SkyrmSigma}).  In its 
small-size limit ($\kappa \rightarrow 0$) the skyrmion is reduced to a hole\cite{Ezawa99JPSJ}.  The 
resultant system is the QH system from which one electron has been removed.  
It corresponds to the loss of the exchange energy $\varepsilon _{X}$ in ferromagnets when 
one electron is removed.  Hence, $\varepsilon _{X}=4\pi J$.  

Consequently, the effective Hamiltonian is given by (\ref{FieldHamil}), 
which is appropriate to analyze phenomena whose scale is larger than the 
magnetic length $\ell _{B}$.  The ground-state exchange energy is given by
\begin{equation}
\langle H_{X}\rangle _{\txg}=-2\pi JN ,
\label{GrounEnergMono}
\end{equation}
together with the spin stiffness (\ref{SpinStiff}).  The effective Hamiltonian 
(\ref{FieldHamil}) describes the spin wave in the QH ferromagnet at $\nu =1$.  The 
spin wave is a Goldstone mode associated with the global O(3) symmetry 
spontaneously broken.  Due to the Zeeman effect the Goldstone mode acquires a 
gap and the coherent length is made finite:  See (\ref{SpinLengt}).

\section{Bilayer Quantum Hall Ferromagnets}

We generalize arguments to analyze electrons in the \LLL\ in BLQH 
systems.  The SU(2) pseudospin structure is introduced by assigning up (down) 
pseudospin to the front (back) layer.  One Landau site contains four electron 
states in the \LLL, which are distinguished by the SU(4) isospin index 
$\sigma =\txf\!\upA ,\txf\!\dnA ,\txb\!\upA ,\txb\!\dnA $.  For instance, $\sigma =\txf\!\upA $ implies that 
the electron is in the front layer and its spin is up.  The group SU(4) is 
generated by the Hermitian, traceless, $4\times 4$ matrices.  There are $(4^{2}-1)$ 
independent matrices.  We take a standard basis\cite{BookGeorgi}, $\lambda _{a}$, 
$a=1,2,\cdots ,15$, normalized as $\text{Tr}(\lambda _{a}\lambda _{b})=2\delta _{ab}$.  They are the 
generalization of the Pauli matrices.  

We decompose the microscopic Coulomb interaction into two terms,
\begin{subequations}\label{BLCouloPM}
\begin{align}
H_{C}^{+}&= {\frac{1}{2}}\int \dxdy V_{+}(\bbox{x}-\bbox{y})\rho (\bbox{x})\rho (\bbox{y}) ,  \label{BLCouloP}\\
H_{C}^{-}&= {\frac{1}{2}}\int \dxdy V_{-}(\bbox{x}-\bbox{y})\Delta \rho (\bbox{x})\Delta \rho (\bbox{y}),  \label{BLCouloM}
\end{align}
\end{subequations}
where $H_{C}^{+}$ depends on the total density $\rho (\bbox{x})$, and $H_{C}^{-}$ on the density 
difference $\Delta \rho (\bbox{x})$ between the front and back layers, 
\begin{equation}
\Delta \rho (\bbox{x})=\rho ^{\txf\upA }(\bbox{x})+\rho ^{\txf\dnA }(\bbox{x})-\rho ^{\txb\upA }(\bbox{x})-\rho ^{\txb\dnA }(\bbox{x}).
\end{equation}
The Coulomb term $H_{C}^{+}$ is invariant under the SU(4) transformation.

The electron field $\psi ^{\sigma }(\bbox{x})$ is expanded as in (\ref{FieldExpan}),
\begin{equation}
\psi ^{\sigma }(\bbox{x}) \equiv  \sum _{i=1}^{N_{\Phi }} c_{\sigma }(i)\F_{i}(\bbox{x}) ,
\label{FieldExpanBL}
\end{equation}
where $c_{\sigma }(i)$ is the annihilation operator of the electron with isospin $\sigma $ at 
site $i$.  Substituting the expansion (\ref{FieldExpanBL}) into the Coulomb term 
(\ref{BLCouloPM}), we extract the direct and exchange Coulomb terms.  Because the 
QH system is robust against density fluctuations, the direct Coulomb term 
arising from the SU(4)-invariant term (\ref{BLCouloP}) is irrelevant as far as 
perturbative fluctuations are concerned.  The direct term from the 
SU(4)-noninvariant term (\ref{BLCouloM}) is
\begin{equation}
H_{\text{cap}} = \varepsilon _{\text{cap}}\sum _{i=1}^{N_{\Phi }} \sS_{z}(i)\sS_{z}(i) ,
\label{CapacEnerg}
\end{equation}
where $\sS_{z}=\sS^{\upA }_{z} + \sS^{\dnA }_{z}$ at each site and
\begin{equation}
\varepsilon _{\text{cap}} = {\frac{e^{2}}{4\pi \varepsilon \ell _{B}}}\sqrt {\frac{\pi}{2}}\biggl(1- e^{d^2/2\ell _{B}^{2}} \bigl\{1-\text{erf}\bigl(d/\sqrt {2}\ell _{B}\bigr)\bigr\}\biggr)
\end{equation}
with the error function $\text{erf}(x)$.  Here, $\sbS^{\upA }(i)$ is the SU(2) 
pseudospin at site $i$ made of the two component spinor $(c_{\txf\upA },c_{\txb\upA })$ as 
in (\ref{ExchaPauli}).  We call the SU(4)-noninvariant Coulomb interaction 
(\ref{BLCouloM}) the capacitance term since $\varepsilon _{\text{cap}}\sS_{z}(i)\sS_{z}(i)$ describes 
the capacitance energy per one Landau site.

We proceed to study the exchange Coulomb interaction.  For this purpose 
we define the SU(4) isospin at each site $i$ by
\begin{equation}
\frS_{a} = (c_{\txf\upA }^{\dagger },c_{\txf\dnA }^{\dagger },c_{\txb\upA }^{\dagger },c_{\txb\dnA }^{\dagger }) {\frac{\lambda _{a}}{2}} 
\begin{pmatrix} c_{\txf\upA }\\ c_{\txf\dnA }\\ c_{\txb\upA }\\ c_{\txb\dnA } \end{pmatrix} .
\label{SinLayer}
\end{equation}
Substituting the expansion (\ref{FieldExpanBL}) into (\ref{BLCouloP}), and using the 
algebraic relation
\begin{equation}
\sum _{\sigma ,\tau } c^{\dagger }_{\sigma }(i)c^{\dagger }_{\tau }(j)c_{\tau }(i)c_{\sigma }(j) = -2\frbS(i)\!\cdot\! \frbS(j) -{\frac{1}{ 4}}n(i)n(j) ,
\label{AppExchaCCCC}
\end{equation}
we obtain the SU(4)-invariant exchange energy as
\begin{equation}
H_{X}^{+} = - 4\sum _{\langle i,j\rangle }J^{+}_{ij} \biggl(\frbS(i)\!\cdot\! \frbS(j) + {\frac{1}{2}}n(i)n(j)\biggr) .
\label{CouloIVBL}
\end{equation}1
The exchange integral $J^{+}_{ij}$ is defined by (\ref{ExchaInteg}) with the use of the 
Coulomb potential $V_{+}(\bbox{x}-\bbox{y})$.   The Hamiltonian (\ref{CouloIVBL}) takes the minimum 
value when the product $\frbS(i)\!\cdot\! \frbS(j)$ takes the maximum value.  It occurs 
for $\frbS(i)=\frbS(j)$, where $\frbS(i)\!\cdot\! \frbS(j)=3/8$ at $\nu =1$.  

It is convenient to decompose the exchange term (\ref{CouloIVBL}) in terms 
of various SU(2) components,
\begin{align}
H^{+}_{X} = &-4\sum _{\langle i,j\rangle }J^{+}_{ij} \biggl(\bbox{S}^{\txf}(i)\!\cdot\! \bbox{S}^{\txf}(j) + \bbox{S}^{\txb}(i)\!\cdot\! \bbox{S}^{\txb}(j)\biggr)_{xy} \notag\\
&-4\sum _{\langle i,j\rangle } J^{+}_{ij}\biggl(\sbS^{\upA }(i)\!\cdot\! \sbS^{\upA }(j)+ \sbS^{\dnA }(i)\!\cdot\! \sbS^{\dnA }(j)\biggr)_{xy}  \notag\\
&-4\sum _{\langle i,j\rangle } J^{+}_{ij}\biggl(\bbox{I}(i)\!\cdot\! \bbox{I}(j)+\widetilde{\bbox{I}}(i)\!\cdot\! \widetilde{\bbox{I}}(j)\biggr)_{xy}  \notag\\
&- 2 \sum _{\langle i,j\rangle }J^{+}_{ij} \biggl(\sum _{\sigma =1}^{4}n^{\sigma }(i)n^{\sigma }(j)\biggr), 
\end{align}
where $\bigl(\bbox{S}^{\txf}(i)\!\cdot\! \bbox{S}^{\txf}(j)\bigr)_{xy}\equiv S^{\txf}_{x}(i)S^{\txf}_{x}(j)+S^{\txf}_{y}(i)S^{\txf}_{y}(j)$, etc., and
\begin{align}
S_{a}^{\txf} = (c_{\txf\upA }^{\dagger },c_{\txf\dnA }^{\dagger }) {\frac{\tau _{a}}{2}} \begin{pmatrix}c_{\txf\upA }\\ c_{\txf\dnA } \end{pmatrix},\quad 
S_{a}^{\txb} = (c_{\txb\upA }^{\dagger },c_{\txb\dnA }^{\dagger }) {\frac{\tau _{a}}{2}} \begin{pmatrix}c_{\txb\upA }\\ c_{\txb\dnA } \end{pmatrix},\notag\\
P_{a}^{\upA } = (c_{\txf\upA }^{\dagger },c_{\txb\upA }^{\dagger }) {\frac{\tau _{a}}{2}} \begin{pmatrix}c_{\txf\upA }\\ c_{\txb\upA } \end{pmatrix},\quad 
P_{a}^{\dnA } = (c_{\txf\dnA }^{\dagger },c_{\txb\dnA }^{\dagger }) {\frac{\tau _{a}}{2}} \begin{pmatrix}c_{\txf\dnA }\\ c_{\txb\dnA } \end{pmatrix},\notag\\
I_{a} = (c_{\txf\upA }^{\dagger },c_{\txb\dnA }^{\dagger }) {\frac{\tau _{a}}{2}} \begin{pmatrix}c_{\txf\upA }\\ c_{\txb\dnA } \end{pmatrix},\quad 
\widetilde{I}_{a} = (c_{\txf\dnA }^{\dagger },c_{\txb\upA }^{\dagger }) {\frac{\tau _{a}}{2}} \begin{pmatrix}c_{\txf\dnA }\\ c_{\txb\upA } \end{pmatrix}.
\end{align}
The exchange energy due to the SU(4)-noninvariant term (\ref{BLCouloM}) is also 
evaluated.  Combining them we obtain
\begin{align}
H_{X} = &-4\sum _{\langle i,j\rangle }J_{ij} \biggl(\bbox{S}^{\txf}(i)\!\cdot\! \bbox{S}^{\txf}(j) + \bbox{S}^{\txb}(i)\!\cdot\! \bbox{S}^{\txb}(j)\biggr)_{xy} \notag\\
&-4\sum _{\langle i,j\rangle } J^{d}_{ij}\biggl(\sbS^{\upA }(i)\!\cdot\! \sbS^{\upA }(j)+ \sbS^{\dnA }(i)\!\cdot\! \sbS^{\dnA }(j)\biggr)_{xy}  \notag\\
&-4\sum _{\langle i,j\rangle } J^{d}_{ij}\biggl(\bbox{I}(i)\!\cdot\! \bbox{I}(j)+\widetilde{\bbox{I}}(i)\!\cdot\! \widetilde{\bbox{I}}(j)\biggr)_{xy} \notag\\
&- 2 \sum _{\langle i,j\rangle }J_{ij} \biggl(\sum _{\sigma =1}^{4}n^{\sigma }(i)n^{\sigma }(j)\biggr),
\label{ExchaHamilBL}
\end{align}
where $J_{ij}^{d}\equiv 2J_{ij}^{+}-J_{ij}$.

The exchange Hamiltonian (\ref{ExchaHamilBL}) is valid at any integer 
filling factor with common exchange integrals $J_{ij}$ and $J_{ij}^{d}$.  It is an 
operator which may act on states possessing various isospins.  We may restrict 
the Hilbert space appropriately at a specific filling factor.  We examine two 
special limits to see that it is reduced to the well-established results at 
$\nu =1$.  First, we apply a large bias voltage and move all electrons to the 
front layer.  The resulting system is dynamically equivalent to the monolayer 
system with spin.  Indeed, by using $S_{z}=(n^{\upA }-n^{\dnA })/2$ and $n=n^{\upA }+n^{\dnA }$, it is easy 
to see that (\ref{ExchaHamilBL}) is reduced to the exchange interaction 
(\ref{HamilFerro}) describing the monolayer QH ferromagnet.  Second, we assume a 
large Zeeman effect so that all spins are forced to be polarized, where the 
system describes the spin-frozen bilayer system.  We now use 
$\sS_{z}=(n^{\txf}-n^{\txb})/2$ and $n=n^{\txf}+n^{\txb}$ to rewrite (\ref{ExchaHamilBL}) as
\begin{align}
&H_{X} \notag\\
&=-4\sum _{\langle i,j\rangle } \biggl(J_{ij} \sS_{z}(i) \sS_{z}(j)+J_{ij}^{d}\bigl(\sS_{x}(i)\sS_{x}(j)+\sS_{y}(i)\sS_{y}(j)\bigr)\biggr) \notag\\
&\phantom{=l}-\sum _{\langle i,j\rangle }J_{ij}n(i)n(j) .
\label{ExchaEnergBL}
\end{align}
By taking the continuum limit as in section \ref{SecContiLimit}, the effective 
Hamiltonian is found to be
\begin{align}
H_{X}^{\eff} &=2J^{d}\int \!d^{2}x\; \bigl(\partial _{i}\sS_{x}(\bbox{x}){\cdot }\partial _{i}\sS_{x}(\bbox{x})+\partial _{i}\sS_{y}(\bbox{x}){\cdot }\partial _{i}\sS_{y}(\bbox{x})\bigr) \notag\\
&+2J\int \!d^{2}x\; \partial _{i}\sS_{z}(\bbox{x}){\cdot }\partial _{i}\sS_{z}(\bbox{x}) ,
\label{FieldBLpre}
\end{align}
where the pseudospin field obeys the normalization condition $\sbS(\bbox{x})^{2}=1/4$ at 
$\nu =1$; the stiffness $J$ is given by (\ref{SpinStiff}) and 
\begin{equation}
J^{d} = {\frac{1}{8}}\rho _{0}\ell _{B}^{4} \int \ddq V(\bbox{q})e^{-|\bbox{q}|d}\bbox{q}^{2}\exp\bigl[-{\frac{\ell _{B}^{2}}{2}}\bbox{q}^{2}\bigr] ,
\label{PpinStiff}
\end{equation}
or
\begin{equation}
{\frac{J^{d}}{J}} = -\sqrt {\frac{2}{\pi}}{\frac{d}{\ell _{B}}}
 +\biggl(1+{\frac{d^{2}}{\ell _{B}^{2}}}\biggr)e^{d^2/2\ell _B^2}\biggl(1-\text{erf}\bigl(d/\sqrt {2}\ell _{B}\bigr)\biggr).
\end{equation}
It agrees with the effective Hamiltonian obtained from the Coulomb energy of 
the pseudospin texture\cite{Moon95B,Ezawa97B}, where $J=\rho _{A}$ and $J^{d}=\rho _{E}$ in their 
notation.

\section{U(1) Gauge Symmetries}

The density imbalance $\sigma _{0}\equiv 2\langle \sS_{z}(i)\rangle $ between the two layers is 
controlled by applying a bias voltage.  It affects the system via the 
interaction term
\begin{equation}
H_{\text{bias}}= - eV_{\text{bias}}\sum _{i} \sS_{z}(i).
\label{BiasHamil}
\end{equation}
The tunneling interaction is
\begin{equation}
H_{T} = - \DSAS \sum _{i}\sS_{x}(i) ,
\label{TunneEnerg}
\end{equation}
where $\DSAS$ is the tunneling gap between the symmetric and antisymmetric 
states.  The total Hamiltonian $H_{\text{tot}}$ is the sum of the exchange term 
(\ref{ExchaHamilBL}) and
\begin{align}
H_{D}= \sum _{i=1}^{N_{\Phi }}\biggl(&-{\ZeeEne} S_{z}(i) + {\varepsilon _{\text{cap}}} \sS_{z}(i)\sS_{z}(i) \notag\\
&-{\DSAS} \sS_{x}(i) - {eV_{\text{bias}}} \sS_{z}(i)\biggr) . 
\end{align}
It consists of the Zeeman term, the capacitance term, the tunneling term and 
the bias term.

Since one electron has four components, we may perform local U(4) 
transformations to the electron field.  However, the Hamiltonian is not 
invariant under most of them.  The symmetry of the direct interaction $H_{D}$ is 
a direct product of two U(1) symmetries, U$^{\upA }$(1)$\otimes $U$^{\dnA }$(1), 
\begin{align}
\begin{pmatrix}\psi ^{\txf\upA }(\bbox{x})\\ \psi ^{\txb\upA }(\bbox{x}) \end{pmatrix} &\longrightarrow  e^{i\alpha (\bbox{x})}\begin{pmatrix}\psi ^{\txf\upA }(\bbox{x})\\ \psi ^{\txb\upA }(\bbox{x}) \end{pmatrix},\notag\\
\begin{pmatrix}\psi ^{\txf\dnA }(\bbox{x})\\ \psi ^{\txb\dnA }(\bbox{x}) \end{pmatrix} &\longrightarrow  e^{i\beta (\bbox{x})}\begin{pmatrix}\psi ^{\txf\dnA }(\bbox{x})\\ \psi ^{\txb\dnA }(\bbox{x}) \end{pmatrix}.
\end{align}
The exchange interaction $H_{X}$ breaks this into a single U(1) symmetry, 
\begin{equation}
\psi ^{\sigma }(\bbox{x}) \longrightarrow  e^{i\alpha (\bbox{x})}\psi ^{\sigma }(\bbox{x}) .
\label{LocalU1}
\end{equation}
This is the exact local symmetry of the total Hamiltonian.  It should be 
emphasized, however, that there is no gapless mode because there is no 
propagating mode associated with it:  Indeed, the kinetic term of the would-be 
phase field, $\psi ^{\sigma \dagger }(\bbox{x})\partial _{k}\psi ^{\sigma }(\bbox{x})$, is absent in the Hamiltonian.  

It is important to recognize\cite{Ghosh01B} that the gauge symmetry 
(\ref{LocalU1}) characterizes the BLQH system.  [See (\ref{GaugeCP}) why we call it the 
gauge symmetry.]  To show this, let us consider a system where the two layers 
are separated sufficiently so that there are no interlayer exchange 
interaction ($J^{d}_{ij}=0$) nor the tunneling interaction ($\DSAS=0$).  Then, the 
total Hamiltonian is invariant under two local transformations, U$^{\txf}$(1) and 
U$^{\txb}$(1), which act on electrons on the two layers independently,
\begin{align}
\begin{pmatrix}\psi ^{\txf\upA }(\bbox{x})\\ \psi ^{\txf\dnA }(\bbox{x}) \end{pmatrix} &\longrightarrow  e^{i\alpha (\bbox{x})}\begin{pmatrix}\psi ^{\txf\upA }(\bbox{x})\\ \psi ^{\txf\dnA }(\bbox{x}) \end{pmatrix},\notag\\
\begin{pmatrix}\psi ^{\txb\upA }(\bbox{x})\\ \psi ^{\txb\dnA }(\bbox{x}) \end{pmatrix} &\longrightarrow  e^{i\beta (\bbox{x})}\begin{pmatrix}\psi ^{\txb\upA }(\bbox{x})\\ \psi ^{\txb\dnA }(\bbox{x}) \end{pmatrix}.
\label{LocalU2}
\end{align}
We may also consider a case without the interlayer exchange interaction 
($J^{d}_{ij}=0$) but with the tunneling interaction ($\DSAS\not=0$).  Then, the 
symmetry (\ref{LocalU2}) is broken into the symmetry (\ref{LocalU1}).  The number of 
U(1) gauge symmetries distinguish various bilayer systems.  We come back to 
this observation to examine the dynamical degrees of freedom in Section 
\ref{SecSUwave}.

\section{Ground-State Energies}

We evaluate the ground-state energy.  Let us consider the case $\nu =1$.  
Unless $\langle \sS^{z}(i)\rangle =\pm 1/2$ electrons are not localized in one of the two layers 
but rather expand over the two layers.  The ground state is the up-spin 
bonding state, which is reduced to the up-spin symmetric state in the balanced 
configuration with $\langle \sS^{z}(i)\rangle =0$.  When $\langle \sS^{z}(i)\rangle =\sigma _{0}/2$, the exchange Coulomb 
energy reads
\begin{align}
\langle H_{X}\rangle _{\nu =1} &=-\sum _{\langle i,j\rangle }\bigl((1+\sigma _{0}^{2})J_{ij}+(1-\sigma _{0}^{2})J^{d}_{ij}\bigr) \notag\\
&=-2\pi \bigl(J^{+}+\sigma _{0}^{2}J^{-}\bigr)N_{\Phi }  ,
\end{align}
where $2J^{\pm }\equiv J\pm J^{d}$.  The ground-state energy is
\begin{align}
\frac{{\langle H\rangle _{\nu =1}}}{ N_{\Phi}}=&-{\frac{1}{2}}(\ZeeEne+\DSAS)+{\frac{1}{4}}\sigma _{0}^{2}\varepsilon _{\text{cap}}-{\frac{1}{2}}\sigma _{0}eV_{\text{bias}} 
 \notag\\
&-2\pi \bigl(J^{+}+\sigma _{0}^{2}J^{-}\bigr) .
\end{align}
In particular, $\langle H\rangle _{\nu =1} = -{\frac{1}{2}}(\ZeeEne+\DSAS)N_{\Phi } -2\pi J^{+}N_{\Phi }$ in the balanced 
configuration ($\sigma _{0}=0$).  

We next consider the case $\nu =2$.  Since two electrons exist in one 
Landau site, we make the composition of (pseudo)spins, ${\frac{1}{2}}\otimes
 {\frac{1}{2}} = 0\oplus 1$.  
We have two types of states within the \LLL; 
(a) three pseudospin-singlet and spin-triplet states (the spin sector),
(b) three pseudospin-triplet and spin-singlet states (the ppin sector).

The spin sector consists of 
$|\txf^{\upA },\txb^{\upA }\rangle $, ${\frac{1}{\sqrt {2}}}\bigl(|\txf^{\upA },\txb^{\dnA }\rangle +|\txf^{\dnA },\txb^{\upA }\rangle \bigr)$, $|\txf^{\dnA },\txb^{\dnA }\rangle $.  
They are the eigenstates of the total Hamiltonian within the sector.  The 
ground state is given by $|\txf^{\upA },\txb^{\upA }\rangle $, and the ground-state energy is 
\begin{equation}
\langle H\rangle ^{\spin}_{\nu =2} = - (\ZeeEne + 4\pi J)N_{\Phi }.
\end{equation}
The state is stable only in the balanced configuration.

The ppin sector consists of 
$|\txf^{\upA },\txf^{\dnA }\rangle $, ${\frac{1}{\sqrt {2}}}\bigl(|\txf^{\upA },\txb^{\dnA }\rangle -|\txf^{\dnA },\txb^{\upA }\rangle \bigr)$, $|\txb^{\upA },\txb^{\dnA }\rangle $.  
Within the sector the total Hamiltonian reads
\begin{equation}
H^{\ppin} = N_{\Phi } \begin{pmatrix}\varepsilon _{\text{cap}}+eV_{\text{bias}} & -\DSAS/\sqrt {2} &0 \\
-\DSAS/\sqrt {2} &0& -\DSAS/\sqrt {2} \\
0 & -\DSAS/\sqrt {2} & \varepsilon _{\text{cap}}-eV_{\text{bias}}\end{pmatrix}
\label{PhaseDiagrMatri}
\end{equation}
apart from a constant exchange energy.  The eigenvalue equation is easily 
solved in the balanced configuration with the zero bias voltage 
($V_{\text{bias}}=0$).  The ground state is given by
\begin{equation}
|\txg\rangle ^{\ppin}_{\nu =2}=-\cos\theta |\txS^{\upA },\txS^{\dnA }\rangle  + \sin\theta |\txA^{\upA },\txA^{\dnA }\rangle  ,
\label{GrounTripleBalan}
\end{equation}
where $\tan\theta =\varepsilon _{\text{cap}}/\bigl(2\DSAS+\sqrt {4\DSAS^{2}+\varepsilon _{\text{cap}}^{2}}\bigr)$.  Here, 
$|\txS^{\upA },\txS^{\dnA }\rangle $ and $|\txA^{\upA },\txA^{\dnA }\rangle $ are the symmetric and antisymmetric 
states.  The ground state is no longer a symmetric state unless 
$\varepsilon _{\text{cap}}=0$ or $d=0$.  A certain amount of the antisymmetric state is 
necessarily mixed due to the capacitance effect.  The ground-state energy is
\begin{equation}
\frac{{\langle H\rangle ^{\ppin}_{\nu =2}}}{ N_{\Phi } } = {\frac{1}{2}}\bigl(\varepsilon _{\text{cap}}-\sqrt {4\DSAS^{2}+\varepsilon _{\text{cap}}^{2}}\bigr) -2\pi (J + 
J^{d}\cos^{2}\theta )
\end{equation}
in the balanced configuration.  The state is stable also in unbalanced 
configurations.

The (pseudo)spin composition at $\nu =3$ reads 
${\frac{1}{2}}\otimes {\frac{1}{2}}\otimes {\frac{1}{2}}    = {\frac{1}{2}}\oplus {\frac{1}{ 2}}\oplus {\frac{3}{ 2}}$, where only the doublet is allowed 
within the \LLL.  The ground state is a pseudospin doublet and spin doublet.  
It is essentially the same as the bilayer state at $\nu =1$.  The ground state 
is
\begin{equation}
|\txg\rangle _{\nu =3}={\frac{1}{ \sqrt {2}}}\bigl(|\txf^{\upA },\txf^{\upA },\txb^{\dnA }\rangle  + |\txf^{\upA },\txb^{\upA },\txb^{\dnA }\rangle \bigr),
\end{equation}
and its energy is 
\begin{equation}
\frac{{\langle H\rangle _{\nu =3}}}{ N_{\Phi }} = -{\frac{1}{ 2}}(\ZeeEne+\DSAS) - \pi (5J + J^{d})
\end{equation}
in the balanced configuration.  The state is stable also in unbalanced 
configuration.

At $\nu =4$ all the Landau sites are filled up.  The ground state is  
pseudospin-singlet and spin-singlet.  The ground state is
\begin{equation}
|\txg\rangle _{\nu =4}=|\txf^{\upA },\txf^{\upA },\txb^{\upA },\txb^{\dnA }\rangle ,
\end{equation}
and its energy is
\begin{equation}
\langle H\rangle _{\nu =4} = -8\pi JN_{\Phi } .
\end{equation}
The state is stable only in the balanced configuration.  

In the SU(4)-invariant limit, where $d\rightarrow 0$ and $J^{d}\rightarrow J$, the exchange 
energy is reduced to a unified formula $\langle H_{X}\rangle =- 2\pi JN$ at $\nu =1 ,2, 3, 4$.  In 
the SU(4)-noninvariant case we note the following intriguing properties.  (A) 
In the ``layer basis'', where we take four independent one-body states 
$|\txf\!\!\upA \rangle $, $|\txf\!\!\dnA \rangle $, $|\txb\!\!\upA \rangle $ and $|\txb\!\!\dnA \rangle $, the exchange 
interaction operates only between the same isospin states, i.e.,
\begin{equation}
\langle H_{X}\rangle =-2\pi JN
\end{equation}
between a pair of $|\txf\!\!\upA \rangle $s, a pair of $|\txf\!\!\dnA \rangle $s, a pair of 
$|\txb\!\!\upA \rangle $s and a pair of $|\txb\!\!\dnA \rangle $s:  All others vanish.  (B) In the 
``SAS'' basis, where we take four independent one-body states $|S\!\!\upA \rangle $, 
$|S\!\!\dnA \rangle $, $|A\!\!\upA \rangle $ and $|A\!\!\dnA \rangle $, we naively expect that the exchange 
interaction operates only between the same isospin states as before, i.e.,
\begin{equation}
\langle H_{X}\rangle =-\pi (J+J^{d})N
\end{equation}
between a pair of $|S\!\!\upA \rangle $s, a pair of $|S\!\!\dnA \rangle $s, a pair of 
\hbox{$|A\!\!\upA \rangle $s}, and a pair of $|A\!\!\dnA \rangle $s: Actually there appears also an 
exchange interaction between different isospin states, i.e.,
\begin{equation}
\langle H_{X}\rangle =-\pi (J-J^{d})N
\end{equation}
between $|S\!\!\upA \rangle $ and $|A\!\!\upA \rangle $, and between $|S\!\!\dnA \rangle $ and $|A\!\!\dnA \rangle $: All 
others vanish.  We recover the naive expectation in the SU(4) invariant limit.  

We explain why the exchange term $J^{d}$ does not appear in the layer 
basis but does in the SAS basis.  It arises for instance from the term 
$\sbS^{\upA }(i)\!\cdot\! \sbS^{\upA }(j)$ in (\ref{ExchaHamilBL}).  We find
\begin{align}
\langle \txf\!\!\upA |\sbS^{\upA }(i)|\txf\!\!\upA \rangle &=
\langle \txf\!\!\dnA |\sbS^{\dnA }(i)|\txf\!\!\dnA \rangle =(0,0,{\frac{1}{2}}),\notag\\
\langle \txb\!\!\upA |\sbS^{\upA }(i)|\txb\!\!\upA \rangle &=
\langle \txb\!\!\dnA |\sbS^{\dnA }(i)|\txb\!\!\dnA \rangle =-(0,0,{\frac{1}{ 2}}),
\end{align}
while
\begin{align}
\langle \txS\!\!\upA |\sbS^{\upA }(i)|\txS\!\!\upA \rangle &=
\langle \txS\!\!\dnA |\sbS^{\dnA }(i)|\txS\!\!\dnA \rangle =({\frac{1}{ 2}},0,0),\notag\\
\langle \txA\!\!\upA |\sbS^{\upA }(i)|\txA\!\!\upA \rangle &=
\langle \txA\!\!\dnA |\sbS^{\dnA }(i)|\txA\!\!\dnA \rangle =-(\frac{1}{2},0,0).
\end{align}
Because only the $x$ and $y$ components contributes to the exchange 
interaction (\ref{ExchaHamilBL}), there is no contribution in the layer basis but 
there is in the SAS basis.  

\section{SU(4) Soft Waves}\label{SecSUwave}

We investigate the SU(4) soft waves at $\nu =1$, which are perturbative 
excitations supported by the exchange interaction.  To identify the dynamical 
degree of freedom we use the composite boson (CB) theory of quantum Hall 
ferromagnets\cite{Ezawa99L,Ezawa99JPSJ} by attaching flux quanta to 
electrons\cite{Girvin87L,Zhang89L,Read89L}.  The CB field $\phi ^{\sigma }(\bbox{x})$ is defined by 
making a singular phase transformation to the electron field $\psi ^{\sigma }(\bbox{x})$,
\begin{equation}
\phi ^{\sigma }(\bbox{x})=e^{-ie\Theta (\bbox{x})}\psi ^{\sigma }(\bbox{x}) , 
\label{DressField}
\end{equation}
where the phase field $\Theta (\bbox{x})$ attaches one flux quantum to each electron via 
the relation, $\varepsilon _{ij}\partial _{i}\partial _{j}\Theta (\bbox{x})=\DiracU \rho (\bbox{x})$.  We then introduce the normalized CB 
field $n^{\sigma }(\bbox{x})$ by
\begin{equation}
\phi ^{\sigma }(\bbox{x})= \phi (\bbox{x})n^{\sigma }(\bbox{x}) , 
\label{SkyrmCB}
\end{equation}
so that the $N$-component field $n^{\sigma }(\bbox{x})$ obeys the constraint
\begin{equation}
\bbox{n}^{\dagger }(\bbox{x}){\!\cdot\! }\bbox{n}(\bbox{x})=\sum _{\sigma }n^{\sigma \dagger }(\bbox{x})n^{\sigma }(\bbox{x})=1 .
\label{ConstCP}
\end{equation}
It follows that $\phi (\bbox{x})=\sqrt {\rho (\bbox{x})}$.  Because the QH system is robust against 
density fluctuations, as far as perturbative fluctuations are concerned, we 
may set $\rho (\bbox{x})=\rho _{0}$, or
\begin{equation}
\phi ^{\sigma }(\bbox{x})=\sqrt {\rho _{0}}n^{\sigma }(\bbox{x}) , 
\label{NoSkyrmCB}
\end{equation}
where $\rho _{0}=N/S$ is the average electron density.  

We count the number of independent fields.  The field $\bbox{n}(\bbox{x})$ consists 
of 4 complex fields, but one real field is eliminated by the constraint 
(\ref{ConstCP}).  Furthermore, the U(1) phase field is not dynamical due to the 
gauge symmetry (\ref{LocalU1}), or
\begin{equation}
n^{\sigma }(\bbox{x})\rightarrow e^{i\alpha (\bbox{x})}n^{\sigma }(\bbox{x}).
\label{LocalCP}
\end{equation}
See also (\ref{ExchaCPx}).  This is the only gauge symmetry in the BLQH system.  
Hence, it contains only 3 independent complex fields.  Such a field is the 
CP$^{3}$ field\cite{DAdda78NPB}.  

A comment is in order.  But for the tunneling interaction and the 
interlayer exchange interaction, the symmetry group is given by (\ref{LocalU2}), or 
\begin{align}
\begin{pmatrix}n^{\txf\upA }(\bbox{x})\\ n^{\txf\dnA }(\bbox{x}) \end{pmatrix} &\longrightarrow  e^{i\alpha (\bbox{x})}\begin{pmatrix}n^{\txf\upA }(\bbox{x})\\ n^{\txf\dnA }(\bbox{x}) \end{pmatrix},\notag\\
\begin{pmatrix}n^{\txb\upA }(\bbox{x})\\ n^{\txb\dnA }(\bbox{x}) \end{pmatrix} &\longrightarrow  e^{i\beta (\bbox{x})}\begin{pmatrix}n^{\txb\upA }(\bbox{x})\\ n^{\txb\dnA }(\bbox{x}) \end{pmatrix}.
\label{LocalCP2}
\end{align}
Because there exist two U(1) gauge symmetries, we have a set of two CP$^{1}$ 
fields rather than one CP$^{3}$ field.

We conclude that the dynamical field is the CP$^{3}$ field in the BLQH 
system due to the exchange interaction and the tunneling interaction.  See 
also Section \ref{SecSkyrm}.  The isospin field $\frbS$ and the CP$^{3}$ field 
$\bbox{n}$ are related by
\begin{equation}
\frS_{a}(\bbox{x})=\bbox{n}^{\dagger }(\bbox{x}){\frac{\lambda _{a}}{2}}\bbox{n}(\bbox{x}) .
\label{IpinDensi}
\end{equation}
Though there are 15 isospin components, only 6 of them are independent.

Let us first analyze the exchange Hamiltonian in the SU(4)-invariant 
limit, where the exchange interaction (\ref{CouloIVBL}) yields a nonlinear sigma 
model
\begin{equation}
H_{X}^{\eff} = 2J^{+} \sum _{a=1}^{15}\int d^{2}x [\partial _{k}\frS_{a}(\bbox{x})]^{2} .
\label{ExchaSUFour}
\end{equation}
The SU(4) isospin field obeys the normalization condition $\frbS(\bbox{x})^{2}=3/8$ at 
$\nu =1$.  

Since the independent fields are the CP$^{3}$ fields $n^{\sigma }(\bbox{x})$, we rewrite 
the exchange Hamiltonian (\ref{ExchaSUFour}) in terms of them.  Let us define 
\begin{equation}
{\cal T}(\bbox{x}) = \sum _{a}T^{a}(\bbox{x}){\frac{\lambda ^{a}}{ 2}} .
\end{equation}
We then have
\begin{equation}
{\cal T}^{\alpha \beta }(\bbox{x}) = -{\frac{1}{2N}}\bigl(\delta ^{\alpha \beta } - N n^{\beta \dagger }(\bbox{x})n^{\alpha }(\bbox{x})\bigr) 
\end{equation}
with $N=4$ for the SU(4) isospin field.  Using this, it is straightforward to 
derive\cite{Eichenherr78NPB} from (\ref{ExchaSUFour}) that
\begin{equation}
H_{X}^{\eff} = 2J^{+} \int d^{2}x \bigl\{(\partial _{j}\bbox{n}^{\dagger }\!\cdot\! \partial _{j}\bbox{n}) - (\bbox{n}^{\dagger }\!\cdot\! \partial _{j}\bbox{n})(\partial _{j}\bbox{n}^{\dagger }\!\cdot\! \bbox{n})\bigr\}.
\label{ExchaCP}
\end{equation}
This Hamiltonian has the U(1) gauge symmetry (\ref{LocalCP}).  To see this more 
explicitly, we rewrite it as\cite{Eichenherr78NPB,DAdda78NPB}
\begin{equation}
H_{X}^{\eff} = 2J^{+} \int d^{2}x (\partial _{j}\bbox{n}^{\dagger }+ iK_{j}\bbox{n}^{\dagger })\!\cdot\! (\partial _{j}\bbox{n}\bbox{ }- iK_{j}\bbox{n}) ,
\label{ExchaCPx}
\end{equation}
with
\begin{equation}
K_{\mu }(\bbox{x})=-i\bbox{n}^{\dagger }(\bbox{x})\partial _{\mu }\bbox{n}(\bbox{x}) .
\label{AuxiaK}
\end{equation}
The Hamiltonian (\ref{ExchaCPx}) is invariant under the gauge tranformation
\begin{equation}
n^{\sigma }(\bbox{x})\rightarrow e^{i\alpha (\bbox{x})}n^{\sigma }(\bbox{x}),\quad  K_{\mu } \rightarrow  K_{\mu } + \partial _{\mu }\alpha (\bbox{x}) .
\label{GaugeCP}
\end{equation}
Here, the field $K_{\mu }$ is not a dynamical field\cite{DAdda78NPB}, since it is an 
auxiliary field defined by (\ref{AuxiaK}).

\begin{figure}[tbp]
\includegraphics*[width=85mm]{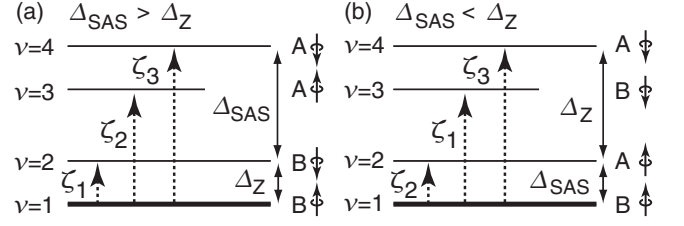}
\renewcommand{\baselinestretch}{1.0}
\caption{
The \LLL\ contains four energy levels corresponding to the two layers and the 
two spin states.  They are represented (a) for $\DSAS>\ZeeEne$ and (b) for 
$\DSAS<\ZeeEne$.  The lowest-energy level consists of up-spin bonding states, 
and is filled at $\nu =1$.  Small fluctuations are Goldstone modes $\zeta _{1}$, $\zeta _{2}$ and 
$\zeta _{3}$.
}\label{BLvaccPS}
\end{figure}

We study small fluctuations of the SU(4) soft waves in the balanced 
configuration with no bias voltage ($V_{\text{bias}}=0$).  The ground state is an 
up-spin symmetric state at $\nu =1$.  It is convenient to use the SAS basis 
rather than the layer basis.  The ground state is given by
\begin{equation}
(n^{\txS\upA },n^{\txS\dnA },n^{\txA\upA },n^{\txA\dnA }) = (1,0,0,0) .
\label{GrounCP}
\end{equation}
We expand the CP$^{3}$ field up to the first order of fluctuation fields 
[Fig.\ref{BLvaccPS}], 
\begin{equation}
(n^{\txS\upA },n^{\txS\dnA },n^{\txA\upA },n^{\txA\dnA }) \simeq  (1,\zeta _{1},\zeta _{2},\zeta _{3}),
\label{IsospFluct}
\end{equation}
where
\begin{equation}
\zeta _{i}(\bbox{x}) = {\frac{1}{ 2}}\bigl(\sigma _{i}(\bbox{x})+i\vartheta _{i}(\bbox{x})\bigr) .
\label{ZetaField}
\end{equation}
They are canonical fields obeying [see (\ref{NoSkyrmCB})]
\begin{equation}
[\zeta _{i}(\bbox{x}),\zeta _{j}^{\dagger }(\bbox{y})] = \rho _{0}^{-1}\delta _{ij}\delta (\bbox{x}-\bbox{y}).
\end{equation}
It is manifest that $\rho _{0}\sigma _{i}(\bbox{x})$ denotes the number density excited from the 
ground state $|\txS\!\!\upA \rangle $ to the $i$-th level designated by (\ref{IsospFluct}).  
The field $\vartheta _{i}(\bbox{x})$ is the conjugate phase variable.

We expand the exchange interaction (\ref{ExchaCPx}) up to the second order,
\begin{align}
\H_{X}^{\eff}&={2J^{+}}\sum _{i=1}^{3}\partial _{k}\zeta ^{\dagger }_{i}(\bbox{x})\partial _{k}\zeta _{i}(\bbox{x}) \notag\\
&={\frac{J^{+}}{ 2}}\sum _{i=1}^{3} \bigl\{(\partial _{k}\sigma _{i})^{2} + (\partial _{k}\vartheta _{i})^{2}\bigr\} .
\label{EffecHamilSpin}
\end{align}
This Hamiltonian describes three Goldstone modes associated with spontaneous 
symmetry breakdown of the SU(4) isospin symmetry.

Actually, the SU(4) symmetry is broken explicitly but softly by various 
direct interactions.  Important SU(2) operators are
\begin{align}
S_{z}&\simeq  -{\frac{1}{ 4}}(\sigma _{1}^{2}+\vartheta _{1}^{2}+\sigma _{3}^{2}+\vartheta _{3}^{2})+ {\frac{1}{ 2}}, \\
P_{z}&= {\frac{1}{2}}\sigma _{2},     \label{SPP}\\
P_{x}&\simeq  -\frac{1}{4}(\sigma _{2}^{2}+\vartheta _{2}^{2}+\sigma _{3}^{2}+\vartheta _{3}^{2})+ {\frac{1}{ 2}}, 
\end{align}
up to the second order of fluctuation fields.  Note that (\ref{SPP}) is an exact 
formula.  Direct interaction terms read
\begin{align}
\H_{C}^{-}&=\frac{\varepsilon _{\text{cap}}{\rho _{0}}}{4}\sigma _{2}^{2},  \\
\H_{Z}&= \frac{\ZeeEne{\rho _{0}}}{4}(\sigma _{1}^{2}+\vartheta _{1}^{2}+\sigma _{3}^{2}+\vartheta _{3}^{2}),  \\
\H_{T}&= \frac{\DSAS{\rho _{0}}}{4}(\sigma _{2}^{2}+\vartheta _{2}^{2}+\sigma _{3}^{2}+\vartheta _{3}^{2}).  
\end{align}
Taking into account the SU(4)-noninvariant exchange interaction as well, we 
find that the effective Hamiltonian is decomposed into three independent modes 
$\H=\H_{\spin}+\H_{\ppin}+\H_{\ipin}$, where
\begin{equation}
\H_{\spin} = {\frac{J^{+}}{ 2}}\bigl\{(\partial _{k}\sigma _{1})^{2} + (\partial _{k}\vartheta _{1})^{2}\bigr\}+ \frac{\ZeeEne \rho _{0}}{ 4} \bigl(\sigma _{1}^{2}+\vartheta _{1}^{2}\bigr) ,
\label{SpinWave}
\end{equation}
and
\begin{align}
\H_{\ppin} =&{\frac{J}{2}}(\partial _{k}\sigma _{2})^{2} + {\frac{J^{d}}{ 2}}(\partial _{k}\vartheta _{2})^{2}+\frac{\varepsilon _{\text{cap}}{\rho_{0}}}{4}\sigma _{2}^{2}  \notag\\
&+ \frac{\DSAS{\rho _{0}}}{4}\bigl(\sigma _{2}^{2}+\vartheta _{2}^{2}\bigr)\bbox{ },
\label{PpinWave}
\end{align}
and
\begin{equation}
\H_{\ipin} = {\frac{J^{+}} {2}}\bigl\{(\partial _{k}\sigma _{3})^{2} + (\partial _{k}\vartheta _{3})^{2}\bigr\} - \frac{(\ZeeEne+\DSAS){\rho _{0}}}{ 4} \bigl(\sigma _{3}^{2}+\vartheta _{3}^{2}\bigr) .
\label{IpinWave}
\end{equation}
They describe three independent soft waves, which are pseudo-Goldstone modes 
by acquiring gaps.  Eqs.(\ref{SpinWave}) and (\ref{PpinWave}) agree with the 
results\cite{Moon95B,Ezawa97B} derived previously for the spin wave and the 
pseudospin wave (which we call the ppin wave), respectively.  The group SU(4) 
is more than SU(2)$\otimes $SU(2).  Eq. (\ref{IpinWave}) is the Hamiltonian obtained newly 
for the SU(4) component missed in the SU(2)$\otimes $SU(2) component, which we call 
the ipin mode.  It is notable that the exchange interactions for the spin and 
ipin modes are solely determined by the SU(4)-invariant Coulomb interaction 
(\ref{BLCouloP}).  The SU(4)-noninvariant Coulomb interaction contributes only to 
the interaction Hamiltonian (\ref{PpinWave}) of the ppin mode.  This is because the 
noninvariant term (\ref{BLCouloM}) involves only the density difference $\Delta \rho $ or the 
ppin mode $\sigma _{2}=2P_{z}$.

The coherence lengths (correlation lengths) are not infinite because 
the soft modes are gapful.  They are
\begin{subequations}
\begin{align}
\xi _{\spin}&=\ell _{B}\sqrt {\frac{4\pi J^{+}}{ \ZeeEne}}, \label{SpinLengt}\\
\xi _{\ipin}&=\ell _{B}\sqrt {\frac{{4\pi J^{+}}}{ \ZeeEne+\DSAS}}. \label{IpinLengt}
\end{align}
\end{subequations}
for the spin and ipin modes. The ground state of the ppin mode (\ref{PpinWave}) is 
a squeezed state\cite{Nakajima97B}, where the coherence lengths are different 
between the conjugate variables $\sigma _{2}$ and $\vartheta _{2}$,  
\begin{subequations}
\begin{align}
\xi ^{\vartheta }_{\ppin} &=\ell _{B}\sqrt {\frac{4\pi J^{d}}{ \DSAS}}, \label{PpinLengtL}\\
\xi ^{\sigma }_{\ppin} &=\ell _{B}\sqrt \frac{{4\pi J}}
{\varepsilon _{\text{cap}}+{ \DSAS}}. \label{PpinLengtS}
\end{align}
\end{subequations}
It is notable that $\xi ^{\vartheta }_{\ppin} $ is very large when $\DSAS$ is very small.  
However, $\xi ^{\sigma }_{\ppin} $ is quite small since $\varepsilon _{\text{cap}}$ is quite large in 
actual samples.

It is important that the bias voltage $V_{\text{bias}}$ couples only with 
the ppin wave.  The ipin wave connects the symmetric state with the 
antisymmetric state and requires the tunneling gap for its excitation, but it 
is insensible to the density difference between the two layers.  This is a 
direct consequence of the formula (\ref{SPP}).  Furthermore, it is easy to check 
that the electromagnetic field couples only with the ppin mode because it does 
not affect the spin.  Consequently, the pseudospin wave is only the one that 
is responsible to the coherent tunneling in the BLQH system.  The mode has 
been argued\cite{Ezawa92IJMPB} to lead to the Josephson effect with charge $e$.

\section{CP$^{3}$ Skyrmions}\label{SecSkyrm}

Provided the Zeeman effect is small enough, charged excitations are skyrmions 
in monolayer QH ferromagnets\cite{Sondhi93B}.  They are topological solitons in 
the O(3) nonlinear sigma model (\ref{FieldHamil}).  It should be emphasized that 
the existence of skyrmions is based on the topological reasoning.  It is 
argued as follows.  The dynamical field of the nonlinear sigma model is the 
O(3) spin field $\bbox{S}(\bbox{x})$.  Since it takes value in the 2-sphere S$^{2}$, the 
topological stability is guaranteed based on the theorem $\pi _{2}(\text{S}^{2})=Z$ 
implying that the second homotopy class of S$^{2}$ is the set of integers 
$Z=\{0,\pm 1,\pm 2,\cdots \}$.  The theorem is rephrased as $\pi _{2}(\text{CP}^{1})=Z$.  We now 
argue that skyrmions arise based on the theorem $\pi _{2}(\text{CP}^{3})=Z$ in the BLQH 
system with the SU(4) coherence.

We consider a generic excitation in SU(N) QH ferromagnets at $\nu =1$.  
Here, $N=2$ in monolayer QH ferromagnets and $N=4$ in BLQH ferromagnets.  It 
can be proved\cite{Ezawa99L,Ezawa99JPSJ} that any excitation confined to the \LLL\ 
is expressed in terms of the CB field (\ref{SkyrmCB}) as
\begin{equation}
\phi ^{\sigma }(\bbox{x})
=\sqrt {\rho (\bbox{x})}n^{\sigma }(\bbox{x})=e^{\a(\bbox{x})}\omega ^{\sigma }(z),
\label{DressCB}
\end{equation}
where $\omega (z)$ is an arbitrary analytic function, and $\a(\bbox{x})$ is an auxiliary 
field obeying
\begin{equation}
\bbox{\nabla }^{2}\a(\bbox{x}) = 2\pi  \bigl(\rho (\bbox{x}) - \rho _{0}\bigr) .
\label{DressOperaE}
\end{equation}
The holomorphicity of $\omega ^{\sigma }(z)$ in (\ref{DressCB}) is a consequence of the 
requirement that the excitation is confined within the \LLL.

We solve (\ref{DressCB}) for the CP$^{N-1}$ field,
\begin{equation}
n^{\sigma }(\bbox{x}) =\frac{{\omega ^{\sigma}(z)}}
 {\sqrt{\sum _{\sigma}|\omega ^{\sigma}(z)|^{2}}} .
\label{SkyrmFormu}
\end{equation}
Substituting (\ref{DressCB}) and (\ref{SkyrmFormu}) into (\ref{DressOperaE}) we find
\begin{equation}
{\frac{1}{4\pi}}\bbox{\nabla }^{2}\ln \rho (\bbox{x}) - \rho (\bbox{x}) + \rho _{0}= J^{0}_{\sky}(\bbox{x}) ,
\label{SolitEq}
\end{equation}
where
\begin{equation}
J_{\sky}^{0}(\bbox{x}) = {\frac{1}{4\pi} }\bbox{\nabla }^{2}\ln\sum _{\sigma }|\omega ^{\sigma }(z)|^{2}.
\label{SemiClassCharg}
\end{equation}
With the aid of the Cauchy-Riemann equation for $\omega (z)$ in (\ref{SkyrmFormu}), this 
is shown\cite{Ezawa99JPSJ} to be the time component of the topological 
current\cite{DAdda78NPB} defined by
\begin{equation}
J^{\mu }_{\sky}(\bbox{x}) = {\frac{1}{2\pi} }\varepsilon ^{\mu \nu \lambda }\partial _{\nu }K_{\lambda }(\bbox{x}) ,
\label{TopolCurre}
\end{equation}
with (\ref{AuxiaK}). The topological charge is given by
\begin{equation}
Q_{\sky}=\int d^{2}x J^{0}_{\sky}(\bbox{x}).
\label{TopolCharg}
\end{equation}
It is conserved trivially, $\partial _{\mu }J^{\mu }_{\sky}(\bbox{x})=0$.  

Eq.(\ref{SkyrmFormu}) is the generic formula for skyrmions\cite{DAdda78NPB}.  
Eq.(\ref{SolitEq}) implies that the density modulation $\delta \rho (\bbox{x})\equiv \rho (\bbox{x})-\rho _{0}$ is induced 
around a skyrmion.  It follows from (\ref{SolitEq}) that 
\begin{equation}
\int d^{2}x [\rho (\bbox{x})-\rho _{0}]= -\int d^{2}x J^{0}_{\sky}(\bbox{x}) = -Q_{\sky}= -1,
\end{equation}
as implies that one skyrmion removes one electron.

The key of the topological stability is whether the skyrmion 
configuration (\ref{DressCB}) can be brought into the ground-state configuration by 
a continuous deformation of the CB field.  First, the CP$^{N-1}$ field 
(\ref{SkyrmFormu}) with $Q_{\sky}\not=0$ cannot be deformed continuously into the 
ground-state value based on the topological theorem $\pi _{2}(\text{CP}^{N-1})=Z$.  
Second, the density $\rho (\bbox{x})$ cannot be deformed continuously into the 
ground-state value  $\rho _{0}$ because in the midstream of this deformation the 
field configuration escapes the \LLL.  Indeed, we have shown that $\rho (\bbox{x})$ 
should obey the soliton equation (\ref{SolitEq}) as far as it is confined within 
the \LLL.  Consequently, skyrmions are stable in QH systems because 
$\pi _{2}(\text{CP}^{N-1})=Z$ and the QH system is robust against density fluctuations.

The topological charge (\ref{TopolCharg}) is determined by the highest power 
of $\omega ^{\sigma }(z)$.  We find $Q_{\sky}=n$ if $\omega ^{\sigma }(z)\rightarrow a^{\sigma }z^{n}$ with $\sum _{\sigma }|a^{\sigma }|^{2}\not=0$.  The 
lightest skyrmion has the topological charge $Q_{\sky}=1$.  It is given by the 
choice of $\omega ^{\sigma }(z) = a^{\sigma }z + b^{\sigma }$ with $\bbox{a}\cdot \bbox{b}=0$ in (\ref{SkyrmFormu}).  The skyrmion 
field (\ref{SkyrmFormu}) transforms under the action of SU(N).  Since it is 
specified by two parameters $\bbox{a}$ and $\bbox{b}$ with $\bbox{a}\cdot \bbox{b}=0$, there are $N(N-1)$ 
skyrmion states apart from the translational degree of freedom.  If the energy 
is solely determined by the nonlinear sigma model (\ref{FieldHamil}) or 
(\ref{ExchaSUFour}), all these states are degenerate with the energy given by 
(\ref{SkyrmEnergX}).  When the skyrmion is required to approach a specific ground 
state asymptotically, the parameter $\bbox{a}$ is fixed, and hence there are $N-1$ 
degenerate skyrmion states.  This is physically reasonable since there exists 
one ground state and $N-1$ excitation states in the \LLL\ 
[Fig.\ref{BLvaccPS}].  

Let us review skyrmions in monolayer QH ferromagnets ($N=2$).  The 
skyrmion is required to approach the spin polarized ground state $\bbox{S}=(0,0,1/2)$ 
asymptotically, and there is no degeneracy since $N-1=1$.  The skyrmion 
configuration (\ref{SkyrmSigma}) is uniquely given by $\bbox{\omega }=(z,\kappa )$ in terms of the 
CP$^{1}$ field.  It gives (\ref{SkyrmSigma}) via $S^{a}={\frac{1}{2}}\bbox{n}^{\dagger }\tau ^{a}\bbox{n}$.

We study skyrmions in BLQH systems ($N=4$).  The skyrmion is required 
to approach the ground state (\ref{GrounCP}) asymptotically, and there are $3$ 
degenerate states since $N-1=3$.  Typical three skyrmions are given by
\begin{align}
\bbox{\omega }_{\spin}&=(z,\kappa ,0,0),  \notag\\
\bbox{\omega }_{\ppin}&=(z,0,\kappa ,0), \notag\\
\bbox{\omega }_{\ipin}&=(z,0,0,\kappa ), 
\end{align}
which we call the spin skyrmion, the ppin skyrmion and the ipin skyrmion, 
respectively.  They are essentially SU(2) skyrmions embedded in the SU(4) 
theory.  

The degeneracy of these three types of skyrmions is resolved by the 
Zeeman effect and the tunneling interaction.  Estimation of their excitation 
energies is straightforward\cite{Ezawa99L} and compared with experimental 
data\cite{Sawada98L,Kumada00L}.  As is obvious in Fig.\ref{BLvaccPS}, it depends 
on a competition between the Zeeman effect and the tunneling interaction 
whether spin skyrmions or ppin skyrmions are excited thermally.

\section{Discussion}

We have derived the Landau-site Hamiltonian (\ref{ExchaHamilBL}) for the 
exchange interaction in BLQH systems.  It is valid at any integer filling 
factor.  A field-theoretical Hamiltonian is constructed from it based on the 
von-Neumann-lattice formulation.  We may use it to analyze phenomena whose 
scale is larger than the magnetic length $\ell _{B}$.  We have analyzed carefully 
BLQH states at $\nu =1$.  The dynamical field is the CP$^{3}$ field because of the 
U(1) gauge symmetry inherent in the system.  We have found that there are 
three soft waves and three skyrmions.  They are excitations from the ground 
state to three excitation levels [Fig.\ref{BLvaccPS}] in the \LLL.  

Though there are three types of skyrmions, only the lightest skyrmions 
are excited thermally.  They are spin skyrmions when the Zeeman gap is small 
enough compared with the tunneling gap, while they are ppin skyrmions when the 
tunneling gap is small enough compared with the Zeeman gap.

It is interesting to apply the present results to BLQH systems at 
$\nu =2$.  In a forthcoming paper we would analyze the predicted canted 
antiferromagnetic phase\cite{Zheng97L,Sarma97L,Demler99L}.  We would also examine 
a prediction that one skyrmion is composed of two skyrmions\cite{Ezawa99L}, which 
seems to have some experimental supports\cite{Sawada98L,Kumada00L}.  

\section*{Acknowledgement}
We would like to thank Hiroshi Ishikawa for various discussions on the 
subject.\par

\end{document}